# Structural modeling of 2019-novel coronavirus (nCoV) spike protein reveals a proteolytically-sensitive activation loop as a distinguishing feature compared to SARS-CoV and related SARS-like coronaviruses

Javier A. Jaimes [1], Nicole M. André [1], Jean K. Millet [2] and Gary R. Whittaker [1]*

[1] Department of Microbiology and Immunology, College of Veterinary Medicine, Cornell University, Ithaca NY 14853 USA

[2] Virologie et Immunologie Moléculaires, INRAE, Université Paris-Saclay, 78352 Jouy-en-Josas, France

* Corresponding author: grw7@cornell.edu

**Abstract**

The 2019 novel coronavirus (2019-nCoV) is currently causing a widespread outbreak centered on Hubei province, China and is a major public health concern. Taxonomically 2019-nCoV is closely related to SARS-CoV and SARS-related bat coronaviruses, and it appears to share a common receptor with SARS-CoV (ACE-2). Here, we perform structural modeling of the 2019-nCoV spike glycoprotein. Our data provide support for the similar receptor utilization between 2019-nCoV and SARS-CoV, despite a relatively low amino acid similarity in the receptor binding module. Compared to SARS-CoV, we identify an extended structural loop containing basic amino acids at the interface of the receptor binding (S1) and fusion (S2) domains, which we predict to be proteolytically-sensitive. We suggest this loop confers fusion activation and entry properties more in line with MERS-CoV and other coronaviruses, and that the presence of this structural loop in 2019-nCoV may affect virus stability and transmission.

**Introduction**

Coronaviruses are zoonotic pathogens that are well known to evolve environmentally and infect many mammalian and avian species (Menachery et al., 2017). These diverse viruses often have effective transmission and immune evasion strategies, especially when outbreaks occur within dense human populations. In the past two decades, coronavirus outbreaks have arisen in human populations around the world, each unique but also with similarities. Severe acute respiratory syndrome coronavirus (SARS-CoV) emerged in 2002 in Guangdong province, China causing an outbreak that spread to 26 countries, with more than 8,000 infections and 774 deaths and a case fatality rate of 9.5% (WHO, 2004). More recently, the ongoing Middle East respiratory syndrome coronavirus (MERS-CoV) outbreak that originated in 2012 in Saudi Arabia has spread to 27 countries with 2,494 infections and 858 deaths, with a case fatality rate of 34.4% (WHO, 2019). The recent surfacing of the novel coronavirus 2019-nCoV (first identified on December 12$^{th}$, 2019) was initially detected in Wuhan, Hubei Province, China, and as of January 30$^{th}$, 2020 has spread globally via travelers and has breached the boundaries of 29 countries (Lu et al., 2020). There has been a quarantine on Wuhan, China since January 23 -24$^{th}$, 2020, along with travel restrictions between China and the rest of the world, in an attempt to prevent a global pandemic (Zhou et al., 2020a). Within the 33 provinces of China, as of January 30$^{th}$, 2020 there have been more than 9,000 suspected cases, with more than 5,900 confirmed cases and 106 fatalities as of January 28th, 2020 with a case fatality rate of 2-3% (Lu et al., 2020). However, the number of cases and deaths have been increasing daily since the outbreak started, and at the time this manuscript was submitted (Feb 9th, 2020), 37,592 cases and 814 fatalities have been reported (Johns Hopkins University, 2019). Similar to SARS-CoV and MERS-CoV, 2019-nCoV infections were observed in family clusters and hospital personnel (Chan et al., 2020; Gralinski and Menachery, 2020; Lu et al., 2020; Phan et al., 2020). The outbreak occurring during the winter is another commonality between SARS-CoV and 2019-nCoV (Zhou et al., 2020a).

Clinical signs associated with 2019-nCoV include pneumonia, fever, dry cough, headache and dyspnea which may progress to respiratory failure and death (Gralinski and Menachery, 2020; Huang et al., 2020; Zhou et al., 2020a). The incubation period for 2019-nCoV seems to be longer than for SARS-CoV and MERS-CoV, which have a mean incubation time of 5 to 7 days leading to challenges in contact tracing (Wu et al., 2020). Preexisting conditions and comorbidities such as hypertension, diabetes, cardiovascular disease or kidney disease affect the severity of pathogenesis attributed to SARS-CoV and MERS-CoV, and thus far similar patterns seem to exist with 2019-nCoV (Gralinski and Menachery, 2020; Huang et al., 2020). SARS-CoV and MERS-CoV seem to exhibit deleterious morbidity and mortality on the elderly population (>60 years of age), with 26 deaths in this age group, and 2019-nCoV is currently portraying a comparable trend (Gralinski and Menachery, 2020). In a study of the first 425 patients in Wuhan, China, the median age of the patients was 59 years; there were no cases in children less than 15 years of age and only a single report of a positive RT-PCR test on sputum in an asymptomatic 10 year old male within a family cluster infection (Chan et al., 2020).

The coronaviruses belong to the *Coronaviridae* family and the *Orthocoronaviridae* subfamily, which is divided in four genera; *Alphacoronavirus, Betacoronavirus, Gammacoronavirus,* and *Deltacoronavirus*. SARS-CoV, MERS-CoV, and 2019-nCoV are all betacoronaviruses, a genus that includes many viruses that infect humans, bats, and other wild animals (ICTV, 2018). Betacoronaviruses have many similarities within the ORF1ab polyprotein and most structural proteins, however, the spike protein and accessory proteins portray significant diversity (Cui et al., 2019). MERS-CoV has maintained a stable genome since its emergence in 2012, unlike other coronaviruses that readily evolve and can undergo notable recombination events (Perlman, 2020).

Alphacoronaviruses and betacoronaviruses have mostly been thought to have originated in bats; this includes SARS-CoV, MERS-CoV and 2019-nCoV, as well as other human coronaviruses such as HCoV-NL63 (Cui et al., 2019; Huynh et al., 2012; Perlman, 2020). Gammacoronaviruses and deltacoronaviruses are reported to have an avian origin, but are known to infect both mammals and avian species (Woo et al., 2012). Human infections of bat-origin viruses typically occur through intermediate hosts; for SARS-CoV these hosts are palm civets (*Paguma larvata*) and racoon dogs (*Nyctereutes procyonoides*), and for MERS-CoV the known host is the dromedary camel (*Camelus dromedarius*) (Cui et al., 2019; Xu et al., 2009). SARS-CoV antibodies were first detected in palm civets and the animal handlers in wet markets (Cui et al., 2019). MERS-CoV is thought to have been circulating for at least 30 years within the dromedary camel population based on retrospective antibody testing of serum from 1983 (Cui et al., 2019). The source of the 2019-nCoV outbreak has been reported to be linked to the Huanan seafood wholesale market in Wuhan, where the CDC confirmed that 43 (22%) of the 198 cases had visited the market (Li et al., 2020). The market sells many species including seafood, birds, snakes, marmots and bats (Gralinski and Menachery, 2020). The market was closed on January 1$^{st}$, 2020 and sampling and decontamination have occurred in order to find the source of the infection. Origination of 2019-nCoV from bats has been strongly supported, but the presumed intermediate host remain to be identified; initial reports that 2019-nCoV had an origin in snakes have not been verified (Gralinski and Menachery, 2020; Zhou et al., 2020a).

The coronavirus spike protein (S) is the primary determinant of viral tropism and is responsible for receptor binding and membrane fusion. It is a large (approx. 180 kDa) glycoprotein that is present on the viral surface as a prominent trimer, and it is composed of two domains, S1 and S2 (Belouzard et al., 2012). The S1 domain mediates receptor binding, and is divided into two sub-domains, with the N-terminal domain often binding sialic acid and the C-domain binding a specific proteinaceous receptor (Hulswit et al., 2016). The receptor for SARS-CoV has been identified as angiotensin converting enzyme 2 (ACE2), similar to what has been recently identified with 2019-nCoV (Cui et al., 2019; Hoffmann et al., 2020; Zhou et al., 2020a). HeLa cells expressing or not human, Chinese horseshoe bat, mouse, civet, and pig ACE2 were infected and 2019-nCoV was able to use all receptors except mouse ACE2 (Zhou et al., 2020a). 2019-nCoV was also found not to use dipeptidyl peptidase 4 (DPP4), the receptor for MERS-CoV (Cui et al., 2019; Hoffmann et al., 2020; Zhou et al., 2020a). Following receptor binding, the S2 domain mediates viral-membrane fusion through the exposure of a highly conserved fusion peptide (Lai et al., 2017;

Madu et al., 2009). The fusion peptide is activated through proteolytic cleavage at a site immediately upstream (S2'), which is common to all coronaviruses. In many (but not all) coronaviruses, additional proteolytic priming occurs at a second site located at the interface of the S1 and S2 domains (S1/S2) (Millet and Whittaker, 2015). The use of proteases in priming and activation, combined with receptor binding and ionic interactions (*e.g.* $H^+$ and $Ca^{2+}$) together control viral stability and transmission, and also control the conformational changes in the S protein that dictate the viral entry process into host cells (Belouzard et al., 2012; Heald-Sargent and Gallagher, 2012; Lai et al., 2017). Specifically, SARS-CoV and MERS-CoV both infect type II pneumocytes in vivo, however they individually infect ciliated bronchial epithelial cells and non-ciliated bronchial epithelial cells receptively (Cui et al., 2019). Similar to SARS-CoV and MERS-CoV, 2019-nCoV can infect *ex vivo* with the same range of cell culture lines, *e.g.* Vero E6, Huh-7 cells (Hoffmann et al., 2020). However primary human airway epithelial cells have been reported to be the preferential cell type for 2019-nCoV (Perlman, 2020; Zhu et al., 2020). Overall, how cell tropism of 2019-nCoV reflects a balance of receptor binding, endosomal environment and protease activation, and the specifics of this mechanisms remain to be determined.

The rapid dissemination and sharing of information during the 2019-nCoV outbreak has surpassed that of both MERS-CoV or SARS-CoV, where the latter virus was only identified after several months and with a genome available a month later (Gralinski and Menachery, 2020). The 2019-nCoV was identified and a genome sequence was available within a month from the initial surfacing of the agent in patients (Gralinski and Menachery, 2020). Initial reports identified that 2019-nCoV contains six major open reading frames in the viral genome and various accessory proteins (Zhou et al., 2020a). The SARS-like virus Bat-CoV RaTG13 was observed to have highly homologous conservation of the genome, with two other bat SARS-like viruses (Bat-SL-CoVZC45 and Bat-SL-CoVZXC21) having 89-97% sequence identity (Gralinski and Menachery, 2020). The S protein of 2019-nCoV was found to be approximately 75% homologous to the SARS-CoV spike (Gralinski and Menachery, 2020; Zhou et al., 2020a).

In this study, we perform bioinformatic analyses and homology structural modeling of 2019-nCoV S, in comparison with closely related viruses. We identify a small structural loop at the S1/S2 interface that contains a short insert containing two arginine residues for 2019-nCoV S. These features are missing from all other SARS-CoV-related viruses, but present in MERS-CoV S and in many other coronaviruses. We discuss the importance of this extended basic loop for S protein-mediated membrane fusion and its implications for viral transmission.

**Results**

Comparison of amino acid identity of the spike (S) protein of 2019-nCoV with human SARS-CoV

To obtain an initial assessment of shared and/or specific features of the 2019-nCoV spike (S) envelope glycoprotein, a protein sequence alignment was performed to compare the sequence of the Wuhan-Hu-1 strain of the novel coronavirus with that of the closely related human SARS-CoV S strain Tor2 sequence (Supplementary Fig. 1). The overall percent protein sequence identity found by the alignment was 76% (Fig. 1A). A breakdown of the functional domains of the S protein, based on the SARS-CoV S sequence, reveals that the S1 receptor-binding domain was less conserved (64% identity) than the S2 fusion domain (90% identity). Within S1, the N-terminal domain (NTD) was found to be less conserved (51% identity) compared to the receptor binding domain (RBD, 74% identity). The relatively high degree of sequence identity for the RBD is consistent with the view that 2019-nCoV, like SARS-CoV, may use ACE2 as its host cell receptor, as for SARS-CoV (Hoffmann et al., 2020; Wan et al., 2020; Zhou et al., 2020a). Interestingly, when the more defined receptor binding motif (RBM) was analyzed (*i.e.* the region of SARS-CoV S containing residues that were shown to directly contact the ACE2 receptor) the identity between the two sequences drops to 50%, in this case hinting at possible differences in binding residues involved in the interaction with the receptor and/or binding affinities (Li et al., 2005; Wan et al., 2020). As expected, within the well-conserved S2 domain, subdomain identities were high for the fusion peptide region (FP, 93% identity), high for the heptad-repeat 1 region (HR1, 88% identity), identical for HR2 (100% identity) and high for both the transmembrane and the C-terminal endodomain (TM, 93% identity and E, 97% identity).

Phylogenetic analysis of 2019-nCoV S with other betacoronaviruses

Early phylogenetic studies on 2019-nCoV genomic sequences revealed that it clustered closely with sequences originating from SARS-like sequences from bats, within lineage B of the *Betacoronavirus* genus. Lineage A groups prototypical coronaviruses such as murine hepatitis virus (MHV) and human coronaviruses HCoV-HKU1 and HCoV-OC43 (Fig. 1B). The other highly pathogenic coronavirus, MERS-CoV is found within lineage C, along with related camel-derived MERS-CoV. Lineage C also groups viruses from bats and other mammals such as hedgehogs. Lineage D contains viral species infecting bats. To gain a better understanding of both shared and specific features of 2019-nCoV S protein, a phylogenetic analysis centered on S protein sequences of representatives of the four *Betacoronavirus* lineages was carried out (Fig. 1B). Fifteen sequences of 2019-nCoV S sequences obtained from NCBI and GISAID from China and various export locations world-wide were analyzed along with representative members of lineages A-D betacoronaviruses. The analysis confirmed that all 2019-nCoV S sequences clustered very closely with bat SARS-like sequences, with the closest matching sequence corresponding to a bat coronavirus (bat-CoV) strain named Bat-SL-RaTG13. Other closely related sequences found were from Bat-SL-CoVZC45 and Bat-SL-CoVZXC21. The sub-clade that groups 2019-nCoV, Bat-SL-RaTG13, Bat-SL-CoVZC45 and Bat-SL-CoVZXC21 is distinct from the one grouping human and civet SARS-CoV along with other related bat SARS-like viruses, such as Bat-SL-LYRa3.

Alignments of RBD and cleavage sites of 2019-nCoV and other bat-CoVs

An S protein sequence alignment focusing on the RBD region of 2019-nCoV, SARS-CoV and bat-SARS-related viruses reveals that the N-terminal half of the RBD is relatively well conserved whereas the C-terminal half, which contains the RBM, exhibits more variations (Fig. 2A). Notably, Bat-SL-CovZC45 and Bat-SLCoVZXC2 both have two deletions of 5 and 14 residues within the RBD. The composition of residues found at the two known coronavirus S cleavage sites was performed using alignment data (Fig. 2B and C). The region around arginine 667 (R667) of SARS-CoV S, the S1/S2 cleavage site aligned well with 2019-nCoV and the bat SARS-related sequences (Belouzard et al., 2009). Notably, an arginine at the position corresponding to SARS-CoV R667 is conserved for the other five sequences analyzed. The alignment shows that 2019-nCoV contains a four amino acid insertion $_{681}$PRRA$_{684}$ that is not found in any other sequences, including the closely related bat-SL-RaTG13 (Fig. 2B). Together with the conserved R685 amino acid found in 2019-nCoV at the putative S1/S2 cleavage site, the insertion introduces a stretch of three basic arginine residues that could potentially be recognized by members of the pro-protein convertase family of proteases (Seidah, 2011; Seidah et al., 2013). This insertion was conserved for all fifteen 2019-nCoV sequences analyzed (Supplementary Fig. 2). Within the *Betacoronavirus* genus, the presence of a basic stretch of residues at the S1/S2 site is found for a number of species from lineages A (HCoV-HKU1, MHV, HCoV-OC43) and C (MERS-CoV, BatCoV-HKU5). The four amino acid insertion feature appears unique among lineage B viruses, as all other species analyzed in the extended alignment, none contained the stretch of basic residues identified in 2019-nCoV S (Supplementary Fig. 2). As expected from previous analyses, the S2' cleavage site, located immediately upstream of the fusion peptide and corresponding to the residue position R797 in the case of SARS-CoV, was strictly conserved for 2019-nCoV and closely related bat SARS-related sequences (Fig. 2C). Of note, the leucine (L) residue found at position 792 of the SARS-CoV sequence is substituted to serine (S) residue for 2019-nCoV S as well as the bat SARS-related sequences. The fusion peptide sequence was found to be well conserved for all sequences analyzed.

2019-nCoV S protein homology structure modeling

To gain a deeper understanding of common and possibly distinguishing features found in 2019-nCoV S protein, homology structure modeling was undertaken. The analysis of modeled proteins provides a powerful tool to identify predicted structural characteristics, that can translate into structure-function changes in of the studied protein. Our laboratory has taken advantage of these tools in the past, for structure-function studies of other CoVs S proteins (Belouzard et al., 2012;

Jaimes et al., 2020; Jaimes and Whittaker, 2018). To perform the modeling, it is first necessary to identify the protein structure to be used as template, which will determine the accuracy of the predicted model. The S protein structure of several CoVs including: *Alphacoronavirus*: HCoV-NL63 and feline coronavirus UU4 (FCoV-UU4); *Betacoronavirus*: HCoV-HKU1, MHV, SARS-CoV and MERS-CoV; *Gammacoronavirus*: infectious bronchitis virus (IBV); and *Deltacoronavirus*: porcine deltacoronavirus (PDCoV) have been reported previously (Kirchdoerfer et al., 2016; Shang et al., 2018; Shang et al., 2017; Walls et al., 2016a; Walls et al., 2016b; Yang et al., 2020; Yuan et al., 2017). Considering that genome and S protein alignments have showed that the 2019-nCoV belongs to the *Betacoronavirus* genus, we focused our analysis on the S structures from viruses belonging to this genus. To select the template structure, the S protein amino acid sequences from four representative betacoronaviruses (HCoV-HKU1, MHV, SARS-CoV and MERS-CoV) were aligned and the solved S protein structures were compared to determine their amino acid identity and the overall structural organization similarities among these proteins (Supplementary Fig. 3A). We observed an average of ~30% identity among the four viral S proteins at the amino acid level, with the exception of HCoV-HKU1 and MHV which share an amino acid identity of 59% at the S protein (Supplementary Fig. 3A). Despite the differences at the amino acid level, the overall structure of the four *Betacoronavirus* S proteins showed a similar folding pattern (Supplementary Fig. 3B), and major differences can only be spotted at specific sections of the functional domains where flexible loops are abundant (e.g. RBD and cleavage sites). Considering this, we built a first set of models for the 2019-nCoV S protein based on each of the above-mentioned structures (Supplementary Fig. 4). Interestingly, we found no major differences at the secondary structures among the 2019-nCoV S protein predicted models depending on the S structure that was used as template for the modeling construction. However, extended flexible loops at the RBD and/or clashes between S monomers at the S2 domain level were observed in the 2019-nCoV S models based on HCoV-HKU1, MHV and MERS-CoV (Supplementary Fig. 4 – first three panels). In contrast, the predicted 2019-nCoV S model based on the SARS-CoV S structure displayed a much better organized folding and no major clashes were observed between the S monomers (Supplementary Fig. 4 – last panel).

As we described previously, the identity between 2019-nCoV and SARS-CoV at the S protein amino acid level was 76%, and the phylogenetic analysis grouped the 2019-nCoV in the lineage A of the *Betacoronavirus* genus, closely relating to SARS-CoV, as well as to other CoVs originated in bats (Fig. 1B). These two considerations, in addition to our preliminary modeling results, suggested SARS-CoV S as the most suitable template for modeling the 2019-nCoV S protein. Taking an alternative approach, the S protein sequence of 2019-nCoV was submitted to two structure homology modeling servers (Phyre 2 http://www.sbg.bio.ic.ac.uk/~phyre2/html/page.cgi?id=index and RaptorX http://raptorx.uchicago.edu). For both cases the structural models with highest homology scores were based on the SARS-CoV S template structure (PDB ID 5X58, data not shown), confirming the choice of using SARS-CoV S as template for generating structural models of 2019-nCoV.

To better compare the predicted structural characteristics of the 2019-nCoV, we also performed homology modeling of four S proteins from Bat-CoVs belonging to lineage B in our phylogenetic analysis, that showed to be closely related to 2019-nCoV. The modeled S proteins from the Bat-CoV: RaTG13, CoVZC45, CoVZXC21 and LYRa3 were compared to the predicted structure of 2019-nCoV S and to the template structure od SARS-CoV (Fig. 3). The amino acid homology of the modeled S proteins in comparison to the template SARS-CoV S was ~71% for all the Bat-CoV S with the exception of the LYRa3 S which shares a homology of 84.69% with the template S. Overall, all the modeled S proteins shared a similar folding pattern in comparison to SARS-CoV S and both, S1 and S2 domains showed a uniform organization (Fig. 3). As expected, differences were mostly observed at the flexible loops forming the 'head" of the S1 domain, specially at the NTD region (RBD region), were most of the amino acid variation was observed (Fig. 2A and 3). The S protein amino acid identity among the Bat-CoV (including 2019-nCoV) ranged between 75.33 – 96.65%, being the RaTG13 and the LYRa3 S proteins the ones with the highest and the lowest identity to 2019-nCoV, respectively. Despite amino acid variability, no major changes in the secondary structures and the overall folding of the proteins were observed among the modeled S structures of these viruses, suggesting a conserved organization for all the S proteins of the lineage B including the 2019-nCoV. Nevertheless, differences at the flexible loops in both domains were observed, and their impact in the 2019-nCoV S protein function must be further studied.

Structural modeling of the predicted RBM 2019-nCoV

It was recently reported that the 2019-nCoV has the potential to bind ACE2 as a receptor to infect the target cell (Hoffmann et al., 2020). This finding appears to agree with previous reports describing the ability of bat-CoVs to successfully bind and use ACE2 as a cellular receptor for infection (Ge et al., 2013; Menachery et al., 2016; Yang et al., 2015). This conservation in the receptor usage among SARS-CoV and SARS-like bat-CoVs, contrasts with the high variability that we observed at the amino acid sequence of the RBM (Fig. 2A). Considering the high variability observed at the amino acid level of this functional region, we compared the predicted RBM structure of the 2019-nCoV and bat-CoVs to the one of SARS-CoV. Interestingly, despite the variability the modeled 2019-nCoV predicted RBM displayed a similar organization to SARS-CoV (Fig. 4 – top panel). This was also observed in the RaTG3 and LYRa3 predicted RBM structures (Fig. 4 – middle left and bottom right panels), suggesting that the RBM organization is well conserved among these viruses. In contrast, the predicted RBM of the CoVZC45 and CoVZXC21 viruses showed a different folding at this region in comparison to SARS-CoV (Fig. 4 – middle right and bottom left panels). These two last viruses showed a 5 and, a 14 amino acid deletions at the RBM sequence (Fig. 2A), which can explain the differential folding in the modeled proteins.

Structural modeling of 2019-nCoV S reveals a proteolytically-sensitive loop

Figure 2B shows a four amino acid insertion $_{681}$PRRA$_{684}$, as well as a conserved R685 at the S1/S2 site of the 2019-nCoV. This insertion, which appears to be common among the lineage B of betacoronaviruses, suggests a differential mechanism of activation for the 2019-nCoV compared to other SARS-CoV and SARS-like Bat-CoV. At the structural level, the S1/S2 site has been shown to be difficult to solve for most CoVs structures, resulting in either incomplete structures (missing the complete S1/S2 site) or structures with an altered (*i.e.* mutated) S1/S2 site (Walls et al., 2016a; Yang et al., 2020; Yuan et al., 2017). Solving the structure of the S1/S2 site was also found to be an issue in the SARS-CoV S structure we used for our modeling analyses. We have previously shown that the S1/S2 site can be modeled in other CoV S proteins and it appears to organize as a flexible exposed loop that extends from the S structure and suggest it could be easily accessible for proteolytic activation (Jaimes et al., 2020).

To better study the S1/S2 site structural organization, we modeled the SARS-CoV S protein based on the S structure of MHV (S1/S2 site mutated in the structure), and MERS-CoV and SARS-CoV (S1/S2 site missing in the structure) to see if the predicted structure of the S1/S2 site was similar in despite the template structure. We observed no differences in the modeled SARS-CoV S protein at the S1/S2 site, predicting an exposed flexible loop in all the three models (data not shown). Based on this, we proceeded to compare the S1/S2 site, as well as other major functional elements of the S2 domain (*i.e.* S2' site and fusion peptide), in the predicted structure in our SARS-CoV, 2019-nCoV and Bat-CoV S models (Fig 5). Remarkably, two features appear to exhibit distinctive characteristics in the 2019-nCoV S model: the fusion peptide, which is predicted to be organized in a more compact conformation for 2019-nCoV S than in SARS-CoV S (Fig 5 – surface models) and the region corresponding to the S1/S2 cleavage site which contains R667 in the case of SARS-CoV (Fig. 5 – S1/S2 alignment box and ribbon models). For SARS-CoV and the bat-CoV proteins, the S1/S2 site forms a short loop that appears flanking closely to the side of the trimeric structure. In the case of 2019-nCoV S, the S1/S2 site is predicted to form an extended loop that protrudes to the exterior of the trimer (Fig. 5). This feature suggests that the S1/S2 loop in 2019-nCoV S could be more exposed for proteolytic processing by host cell proteases. As mentioned before, solving structure of the S1/S2 site appears to present difficulties for most of the reported CoV S structures (Fig. 6 – top panel). However, the exposed loop feature has been demonstrated in both modeled and cryo-EM CoV S structures with similar amino acid sequences at the S1/S2 site (*i.e.* FCoV and IBV, respectively) (Fig. 6 – top panel). Interestingly, FCoV viruses do not always display a S1/S2 site (Fig. 6 – top panel), which results in distinct cell entry mechanisms. We also performed an analysis of the S2' site of the 2019-nCoV in comparison to SARS-CoV and bat-CoV S proteins. As expected, differences in the modeled S2' site structure were not predicted in any of the studied spikes (Fig. 5 – S2' ribbon models). This agrees with the fact that the S2' site appears to be conserved in the studied sequences (Fig. 5 – S2' alignment box) and as we described previously, the SARS-CoV functional R797 residue at the cleavage position 1 (P1). as well as the serine at position 798 (cleavage position P1') are conserved in the 2019-nCoV and among the compared bat-CoVs. This feature also appears to be conserved in other CoVs (Fig. 6 – bottom panel), however, mutations in the residues immediately upstream the SARS-CoV R797 residue (or equivalent in each virus), have been shown to result in changes in the proteolytical requirements in other CoVs (Millet and Whittaker, 2015). We observed mutations L to S and T to S, which were located upstream the P1 arginine at positions P3 and P6 in the S2' site of the 2019-nCoV in comparison to SARS-CoV. These mutations were not predicted to alter the structure of the S2' in the 2019-nCoV (Fig. 5 – S2' ribbon models).

**Discussion**

In this study, we show the presence of a distinct insert in the S1/S2 priming loop of 2019-nCoV S, which is not shared with SARS-CoV or any SARS-related viruses. The significance of this is yet to be explored experimentally, but we consider it may fundamentally change the entry pathway of 2019-nCoV compared to other known viruses in *Betacoronavirus* lineage B. The presence of the extended S1/S2 priming loop containing paired basic residues predicts that 2019-nCoV S would most likely be cleaved during virus assembly and delivery to the cell surface by Golgi-resident proprotein convertases such as furin. Indeed, analysis of Western blots of VSV-pseudoparticles containing 2019-nCoV S have shown the presence of cleaved S, in contrast to pseudoparticles containing SARS-CoV S (Hoffmann et al., 2020). In the case of MERS-CoV, but not SARS-CoV, it is known that priming of S by "pre-cleavage" occurs at the S1/S2 site, giving 2019-nCoV cleavage activation properties more in line with MERS-CoV than SARS-CoV (Belouzard et al., 2009; Hoffmann et al., 2020; Kleine-Weber et al., 2018; Millet and Whittaker, 2014). 2019-nCoV is currently believed to be highly SARS-CoV like with respect to its receptor binding, and the modeling studies reported here are broadly in line with this finding, despite the relatively low amino acid identity in the RBM. However it is important to remember that changes in protease usage may allow coronaviruses to undergo receptor-independent entry (virus-cell fusion) as well as affect syncytia formation (cell-cell fusion) and tissue pathology (Gallagher et al., 1992; Menachery et al., 2019; Phillips et al., 2017).

To our knowledge, the enlarged priming loop of 2019-nCov is unique among the viruses in *Betacoronavirus* lineage C. The presence of a distinct insert containing paired basic residues in the S1/S2 priming loop is common in many coronaviruses in *Betacoronavirus* lineage C (*e.g.* MERS-CoV), as well as in lineage A (*e.g.* mouse hepatitis virus, MHV) and lineage d, and is universally found in G*ammacoronavirus* S (*e.g.* IBV) (Shang et al., 2018). It is noticeably absent in most *Alphacoronaviruses*, with the clear exception of type I canine and feline coronaviruses (Jaimes et al., 2020; Millet and Whittaker, 2015). One feature of the distinct insert for of 2019-nCoV that warrants attention relates to potential changes as the virus evolves. An equivalent loop is present in influenza HA (in this case adjacent to the fusion peptide), and insertions of basic residues into the loop are a primary marker of conversion from low pathogencity to highly pathogenic avian influenza virus (*e.g.* H5N1) (Chen et al., 1998). In coronaviruses, such loop modifications are known to affect MHV pathogenesis, and modulate neurovirulence and neuroinvasiveness of HCoV-OC43 (Frana et al., 1985; Le Coupanec et al., 2015). The FCoV is another example where S1/S2 loop modifications appear to lead directly to changes in viral pathogenesis (André et al., 2019; Jaimes et al., 2020; Licitra et al., 2013).

At present 2019-nCoV is behaving in a distinct manner compared to SARS-CoV. We believe our findings are of special importance considering that the available data suggests ACE2 as a suitable cellular receptor for 2019-nCoV entry (Huang and Herrmann, 2020; Letko and Munster, 2020). In our modeling analysis, we observed that the RBM of the 2019-nCoV predicted a similar organization as SARS-CoV and that deletions at this RBM region in other bat-CoVs are reported to not impact its ability to bind ACE2 (Ge et al., 2013; Menachery et al., 2016; Yang et al., 2015). This suggests that instead of receptor binding, the S1/S2 loop is a distinctive feature relevant to 2019-nCoV pathogenesis and marks a unique similarity to MERS-CoV. We would predict that distinct insert in 2019-nCoV S would give the virus biological properties more in line with MERS-CoV and not SARS-CoV, especially with regard to its cell entry pathway. However, it may also impact virus spread and transmission. While many epidemiological features of 2019-nCoV still need to be resolved, there are many features of transmission that align more with MERS-CoV then SARS-CoV. The reproductive number (R0) of 2019-nCoV (approx. 3-4), is higher than SARS-CoV (approx. 2-3) and while MERS-CoV has a low R0 in humans (<1), it is high in camels and in outbreak situations (>3) (Choi et al., 2018; Dighe et al., 2019; Lipsitch et al., 2003; Liu et al., 2020; Zhou et al., 2020b). One notable feature of the S protein S1/S2 cleavage site was first observed during the purification of the MHV S protein for structural analysis (Walls et al., 2016a). MHV with an intact cleavage loop was unstable when expressed, and so we consider that the S1/S2 loop controls virus stability, likely via access to the down-stream S2' site that regulates fusion peptide exposure and activity. As such it will interesting to monitor the effects of S1/S2 loop insertions and proteolytic cleavability in the context of virus transmission, in addition to virus entry and pathogenesis.

**Materials and methods:**

Sequences:

Amino acid sequences of the S protein used in the phylogenetic analysis were obtained from GISAID and NCBI GenBank. GISAID accession numbers (in parenthesis) from which whole genome sequences were obtained were: 2019-nCoV-Foshan-EPI_ISL_406535 (EPI_ISL_406535), 2019-nCoV-Foshan-EPI-ISL-406534 (EPI-ISL-406534), 2019-nCoV-France-EPI_ISL_406596 (EPI_ISL_406596), 2019-nCoV-Guangdong-EPI_ISL_406538 (EPI_ISL_406538), 2019-nCoV-Guangzhou-EPI_ISL_406533 (EPI_ISL_406533), 2019-nCoV-Nonthaburi-EPI_ISL_403962 (EPI_ISL_403962), 2019-nCoV-Shenzhen-EPI_ISL_405839 (EPI_ISL_405839), 2019-nCoV-Taiwan-EPI_ISL_406031 (EPI_ISL_406031), 2019-nCoV-USA-AZ1-ISL_406223 (EPI_ISL_406223), 2019-nCoV-USA-CA1-ISL_406034 (EPI_ISL_406034), 2019-nCoV-USA-IL1-EPI_ISL_404253 (EPI_ISL_404253), 2019-nCoV-USA-WA1-EPI_ISL_404895 (EPI_ISL_404895), 2019-nCoV-Wuhan-WIV06-EPI_ISL_402129 (EPI_ISL_402129), 2019-nCoV-Zhejiang-EPI_ISL_404228

(EPI_ISL_404228), Bat-SL-RaTG13 (EPI_ISL_402131). GenBank accession numbers (in parenthesis) from which whole genome or S gene sequences were obtained were: 2019-nCoV-Wuhan-Hu1 (MN908947.3), Bat-SL-CoVZC45 (MG772933.1), Bat-SL-CoVZXC21 (MG772934.1), Bat-SL-LYRa3 (KF569997.1), BatCoV/133 (DQ648794.1), BatCoV-GCCDC1 (NC_030886.1), BatCoV-HKU4-1 (EF065505.1), BatCoV-HKU5-1 (EF065509.1), BatCoV-HKU9 (NC_009021.1), BatCoV-Neo/PML-PHE1/RSA (KC869678.4), BatCoV-SC2013 (KC869678.4), Bat-CoV-BM48-31 (NC_014470.1), Bat-SL-HKU3-1 (DQ022305.2), Bat-SL-LYRa11 (KF569996.1), Bat-SL-Rf1 (DQ412042.1), Bat-SL-Rs4231 (KY417146.1), Bat-SL-Rs4255 (KY417149.1), Bat-SL-Rs4874 (KY417150.1), Bat-SL-RS672 (FJ588686.1), Bat-SL-WIV1 (KC881007.1), BtRs-BetaCoV/YN2018C (MK211377.1), BtRs-BetaCoV/YN2018D (MK211378.1), camMERS-CoV-HKFU-HKU-13 (KJ650295.1), camMERS-CoV-HKU23 (KF906251.1), camMERS-CoV-KSA-505 (KJ713295.1), camMERS-CoV-NRCE-HKU205 (KJ477102.1), camMERS-CoV-NRCE-HKU270 (KJ477103.2), CivSARS-CoV-SZ3 (P59594.1), HCoV-229E (NC_002645.1), HCoV-HKU1 (AY597011.2), HCoV-OC43 (KF963244.1), Hedgehog-CoV/VMC/DEU (KC545383.1), hMERS-CoV-EMC/2012 (JX869059.2), hMERS-CoV-England-1 (KC164505.2), hMERS-CoV-Jordan-N3 (KC776174.1), hSARS-CoV-BJ01 (AY278488.2), hSARS-CoV-GZ02 (AY390556.1), hSARS-CoV-HKU39849 (JN854286.1), hSARS-CoV-Tor2 (NC_004718.3), MHV-A59 (M18379.1).

For S protein modeling, amino acid sequences of SARS-CoV Urbani (AAP13441.1), 2019-nCoV-Wuhan-Hu1 (MN908947.3), , Bat-SL-CoVZC45 (MG772933.1), Bat-SL-CoVZXC21 (MG772934.1), Bat-SL-LYRa3 (KF569997.1) and FCoV WSU-79-1683 (JN634064.1) were obtained from the NCBI

GenBank, and Bat-SL-RaTG13 (EPI_ISL_402131) was obtained from GISAID data base. Amino acid sequence of the FCoV-TN406 S was provided by Prof. Susan Baker (Loyola University Chicago).

Amino acid alignments and phylogenetic trees:

Sequences alignments were performed on coronavirus S protein sequences using MAFFT v7.388 (Katoh et al., 2002; Katoh and Standley, 2013). A Maximum-Likelihood phylogenetic tree based on the S protein alignment was generated using PhyML (Guindon et al., 2010). Numbers at nodes indicate bootstrap support (100 bootstraps). Sequence alignment display and formatting was performed using Geneious R10 (Biomatters) and phylogenetic tree display and formatting was performed using FigTree 1.4.3 (http://tree.bio.ed.ac.uk/).

S protein modeling:

Beta-CoV S protein structures and amino acid sequences were obtained from the RCSB Protein Data Base: HCoV-HKU1 (PDB# 5I08), MHV (PDB# 3JCL), MERS-CoV (PDB# 6Q05), SARS-CoV (PDB# 5X58), FCoV-UU4 (PDB# 6JX7), IBV-M41 (PDB# 6CV0) and HCoV-NL63 (PDB# 5SZS). Pairwise amino acid alignments between each of the Beta-CoV and the 2019-nCoV were performed using Geneious Prime® (v.2019.2.3. Biomatters Ltd.) and exported as *.FASTA file extension for further application. S protein models were built using UCSF Chimera (v.1.14, University of California) through modeler homology tool of the Modeller extension (v.9.23, University of California) and edited using PyMOL (v.2.0.7, Schrodinger LLC.). 2019-nCoV S models built was based on HCoV-HKU1, MHV, MERS-CoV and SARS-CoV S structures. Models for the Bat-CoV RaTG13, Bat-CoV CoVZC45, Bat-CoV CoVZXC21, Bat-CoV LYRa3 and SARS-CoV were built based on SARS-CoV S structure. Finally, FCoV-TN406 and FCoV WSU-78-1683 S models were built based on HCoV-NL63 S structure. Additional SARS-CoV S models based on MHV and MERS-CoV S structures were built to validate our modeling approach (data not shown).

**Acknowledgements**

We thank all member of the Whittaker and Daniel labs at Cornell University for comments and discussion, and Joshua Chappie for invaluable help with structural modeling. Work in the author's laboratory is supported by the National Institutes of Health (research grant R01AI35270).

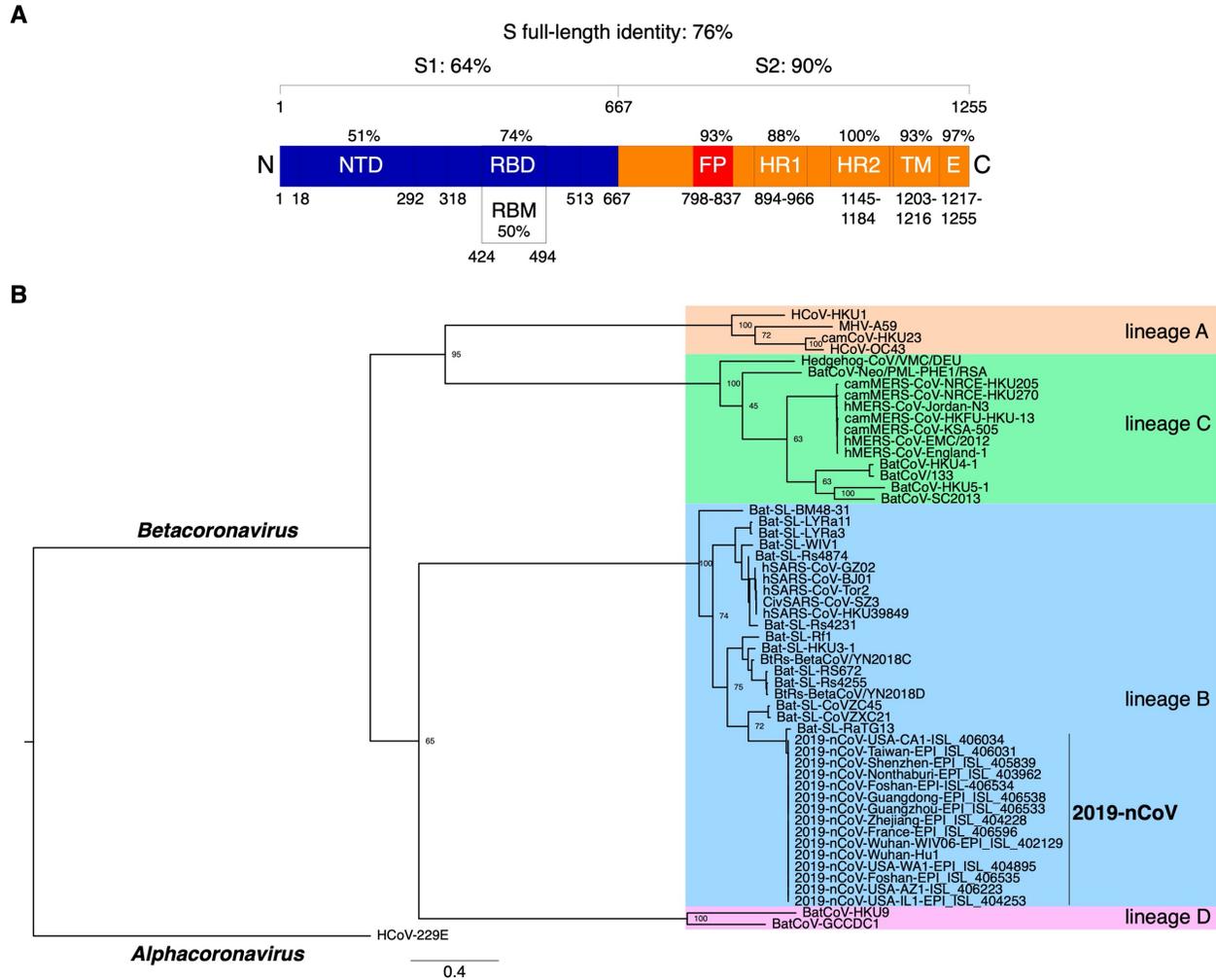

**Figure 1. Comparative analyses of 2019-nCoV S protein sequence. A.** Protein sequence identities between 2019-nCoV S with SARS-CoV S. The S protein sequences were aligned and the sequence identities obtained for the full-length and domains/subdomains are shown on the S protein diagram. Amino acid delineations of domains and subdomains are based on the SARS-CoV S protein. **B.** Phylogenetic analysis of 2019-nCoV S protein. The S protein sequence of 15 isolates of 2019-nCoV were aligned with representatives of all four *Betacoronavirus* lineages. A Maximum-Likelihood tree was generated based on the alignment. The tree was rooted using the alphacoronavirus HCoV-229E S sequence. Number at nodes indicate bootstrap support (100 replicates). Accession numbers of sequences used in the analyses are found in the methods section.

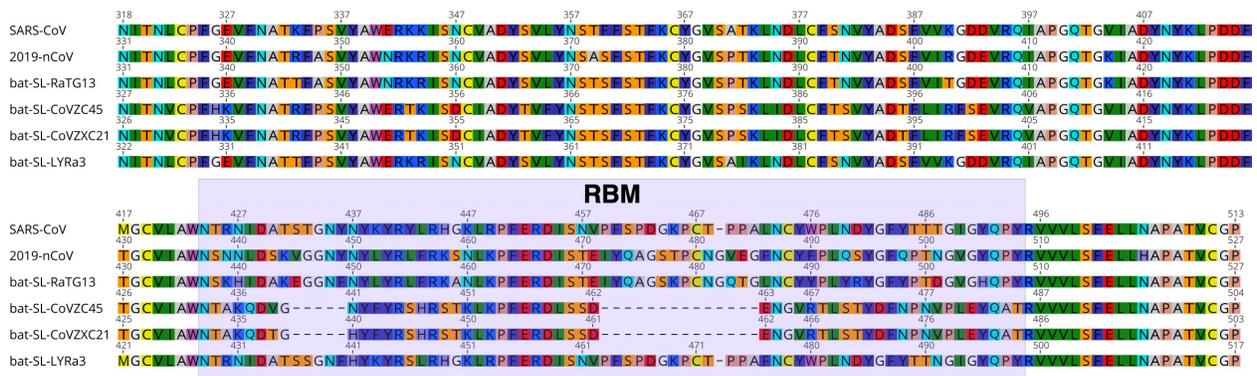
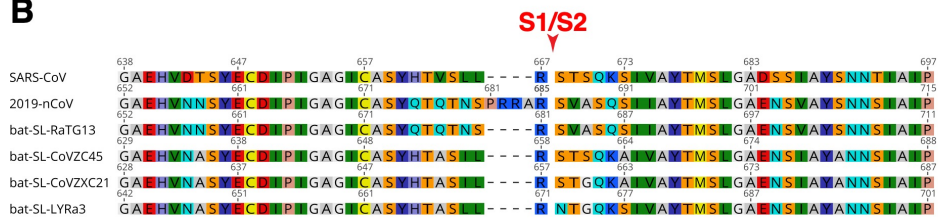
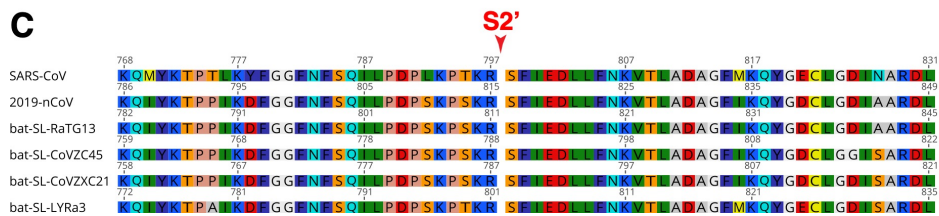

**Figure 2. Sequence alignments of S protein regions of 2019-nCoV with closely related species.** S protein sequences from 2019-nCoV, SARS-CoV, and closely related bat coronaviruses, bat-SL-RaTG13, bat-SL-CoVZC45, bat-SL-CoVZXC21, bat-SL-LYRa3 were aligned. The regions corresponding to the receptor binding domain (RBD, **A.**), the S1/S2 cleavage site (red arrow, **B.**) and S2' cleavage site (red arrow, **C.**) are shown. RBM: receptor binding motif. Accession numbers of sequences used in the analyses are found in the methods section.

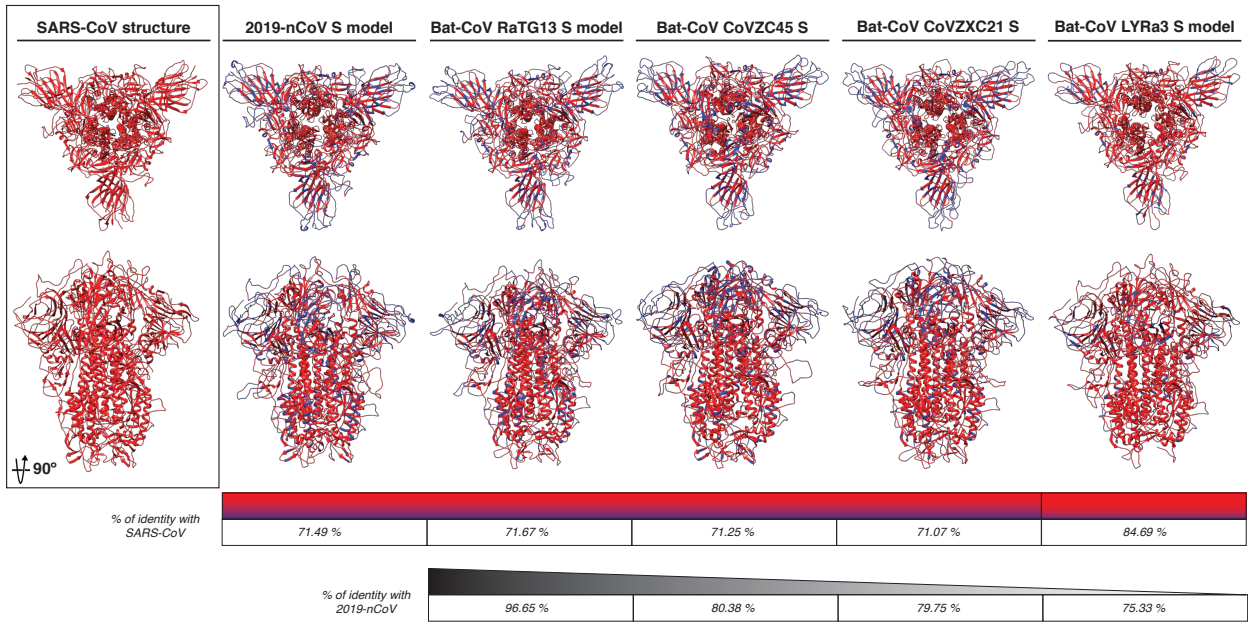

**Figure 3. 2019-nCoV and bat-CoVs S protein models.** The modeled S protein of 2019-nCoV, RaTG13, CoVZC45, CoVCZXC21 and LYRa3 are compared to SARS-CoV S structure. The amino acid homology between the modeled proteins and SARS-CoV S is noted in red and amino acid differences in blue in both models and identity scale. The S amino acid identity between 2019-nCoV and bat-CoVs is also noted (black identity scale).

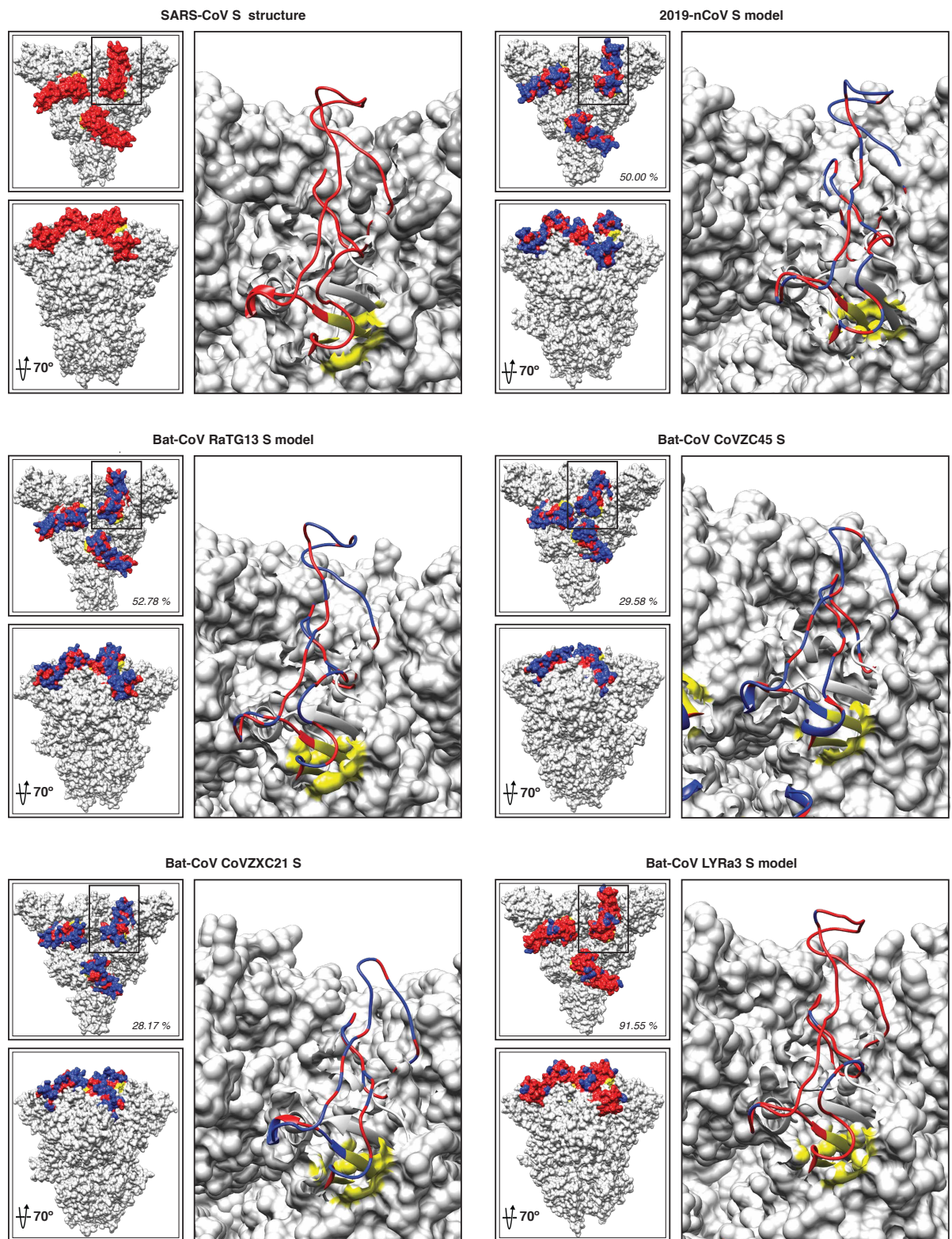

**Figure 4. 2019-nCoV and bat-CoVs modeled RBM.** Surface view of SARS-CoV S structure and 2019-nCoV, RaTG13, CoVZC45, CoVCZXC21 and LYRa3 S models. SARS-CoV RBM (red) and flanking residues (yellow) are noted. RBM in the modeled structures is also noted according to their amino acid homology (red) and differences (blue) to SARS-CoV.

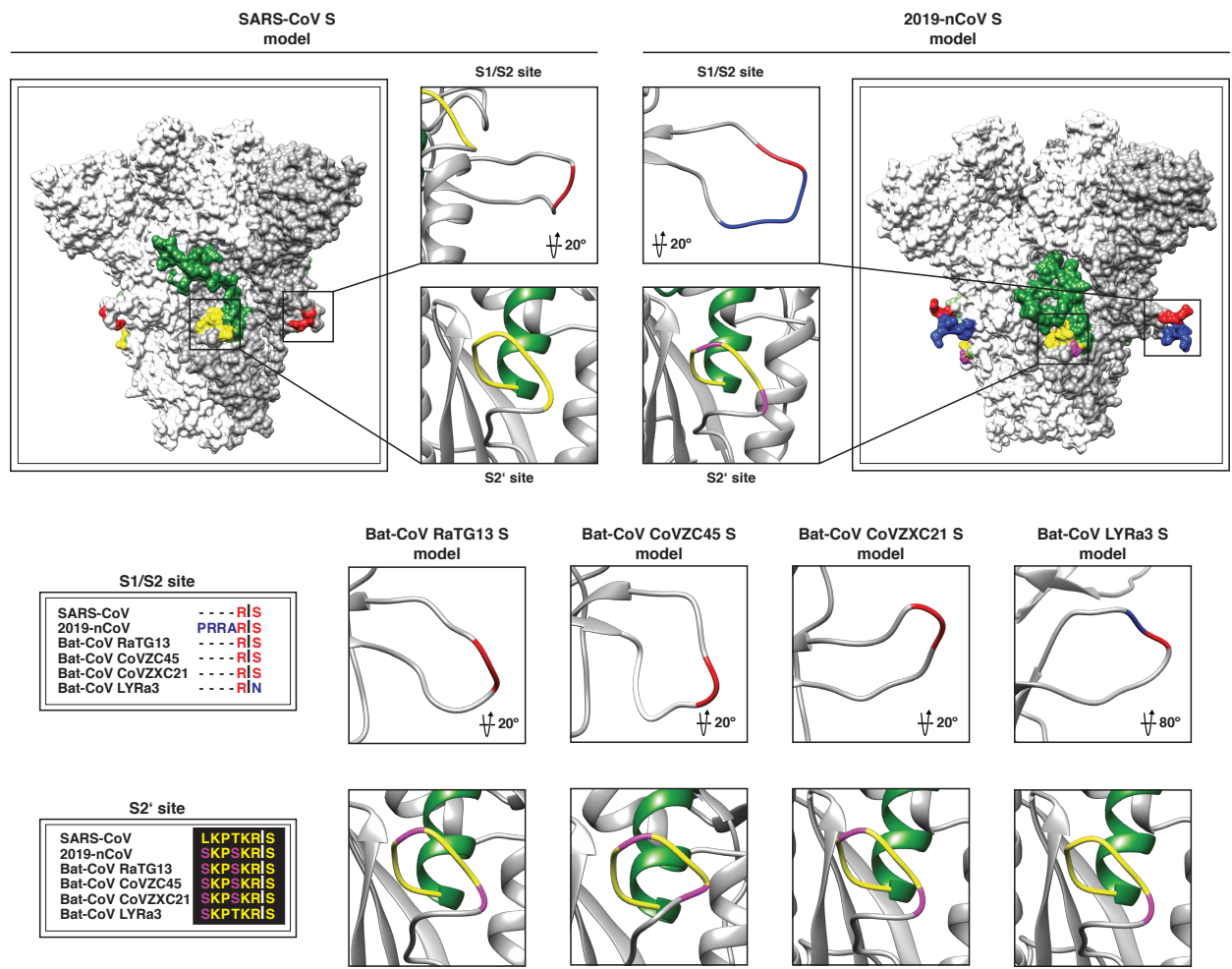

**Figure 5. 2019-nCoV S1//S2 and S2' activation sites.** The S1/S2 and S2' activation sites of SARS-CoV and 2019-nCoV S models are shown in surface and ribbon views. S1/S2 and S2' sites of bat-CoVs are shown in ribbon view. Amino acid homology to SARS-CoV is noted as follows: S1/S2 site: homology (red) and differences (blue); S2' site: homology (yellow) and differences (magenta). Amino acid alignments of the S1/S2 and S2' sites are shown, and homology is also noted.

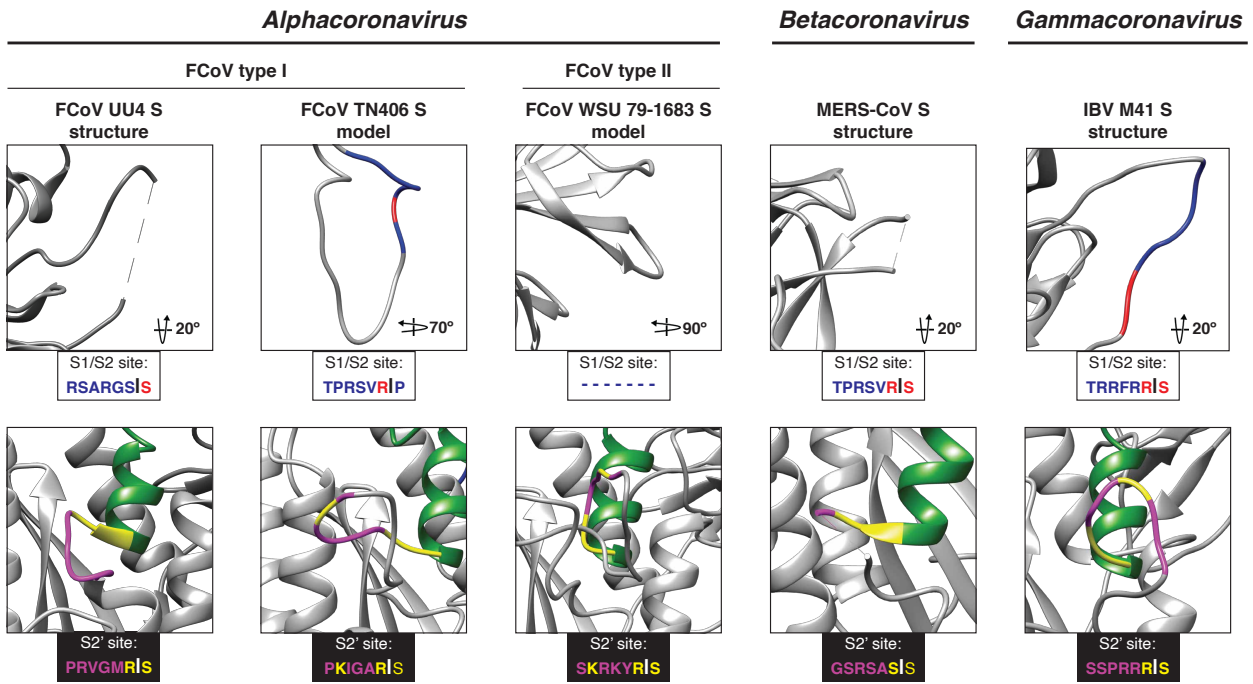

**Figure 6. CoVs S1/S2 and S2' site.** The S1/S2 and S2' activation sites of FCoV, MERS-CoV and IBV. S models are shown ribbon views. Amino acid homology to SARS-CoV is noted as follows: S1/S2 site: homology (red) and differences (blue); S2' site: homology (yellow) and differences (magenta). Amino acid sequences of the S1/S2 and S2' sites are shown.

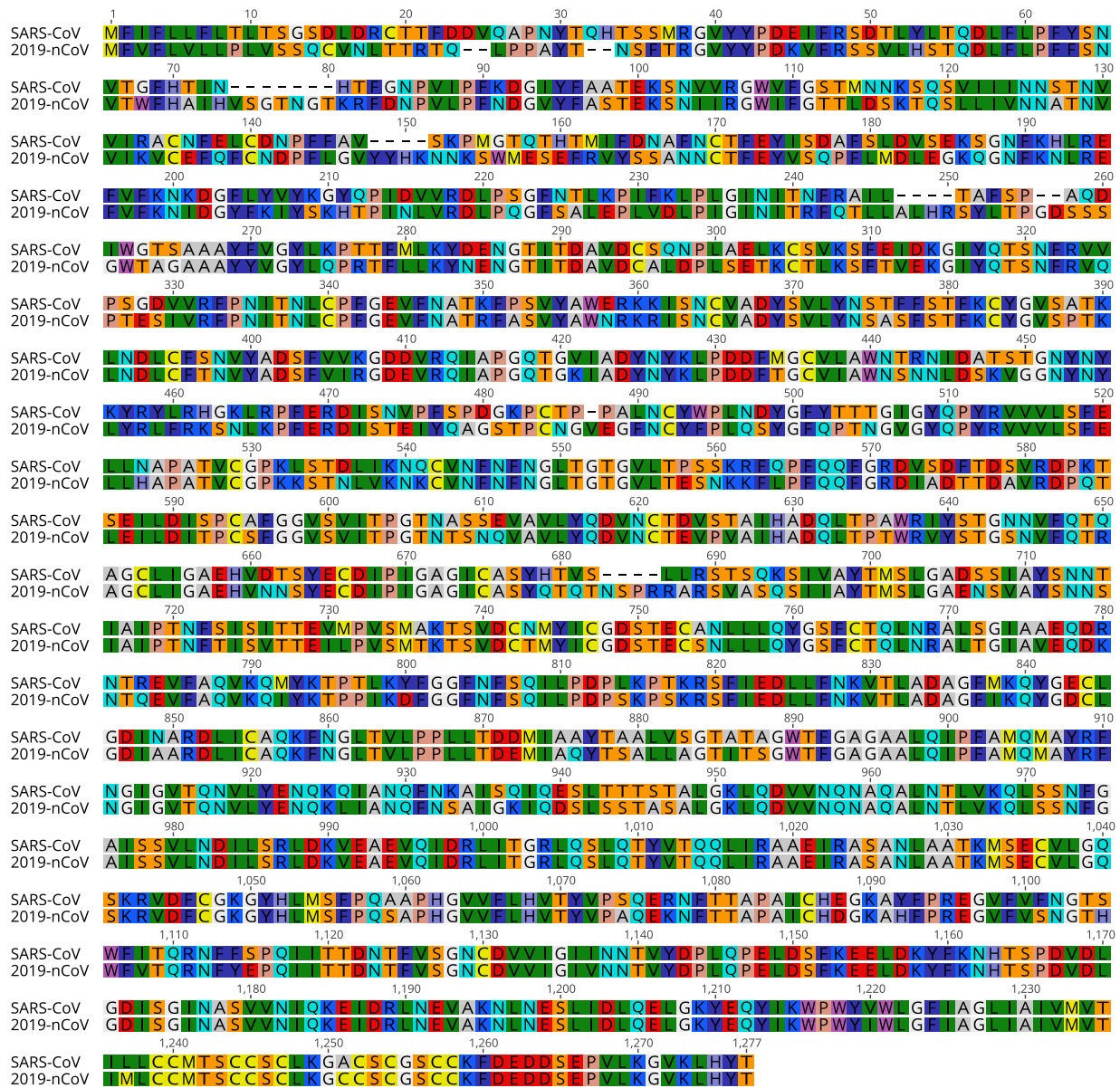

**Supplementary figure 1. Sequence alignment of the S protein of 2019-nCoV with SARS-CoV.** The sequences of the S proteins of 2019-nCoV and SARS-CoV were aligned and the full-length alignment is shown. Accession numbers of sequences used in the analysis are found in the methods section.

**Supplementary figure 2. Sequence alignment of the S1/S2 region of 2019-nCoV with**

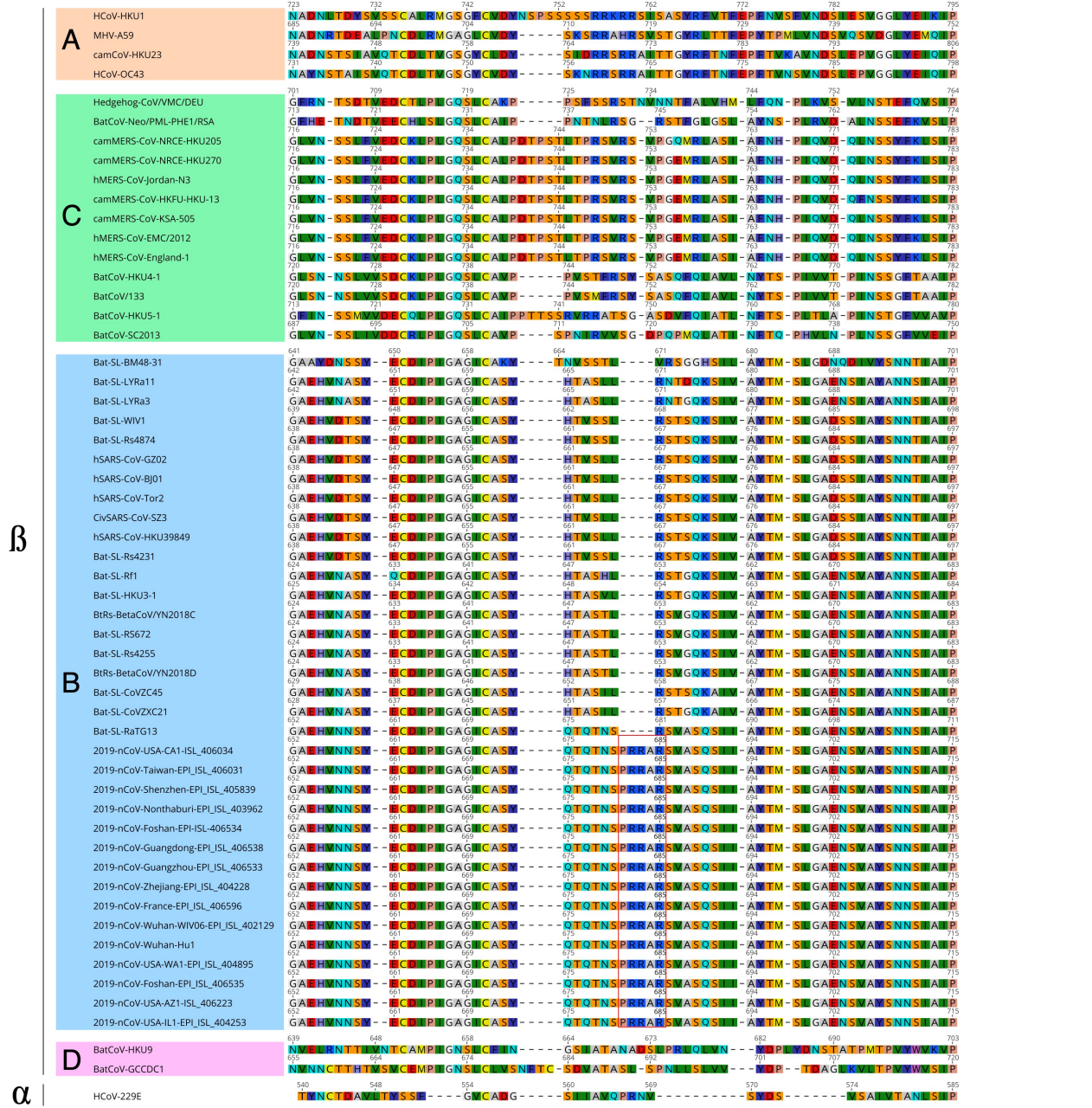

**representatives of betacoronavirus lineages.** The S protein sequence of 15 isolates of 2019-nCoV were aligned with representatives of all four *Betacoronavirus* lineages (lineages are denoted by the letters A, B, C, and D). The sequence for the *Alphacoronavirus* HCoV-229E is also included in the alignment. Displayed is the region corresponding to the S1/S2 cleavage site. Red border indicates the insert of basic arginines identified for 2019-nCoV sequences. α: *Alphacoronavirus*; β: *Betacoronavirus*. Accession numbers of sequences used in the analyses are found in the methods section.

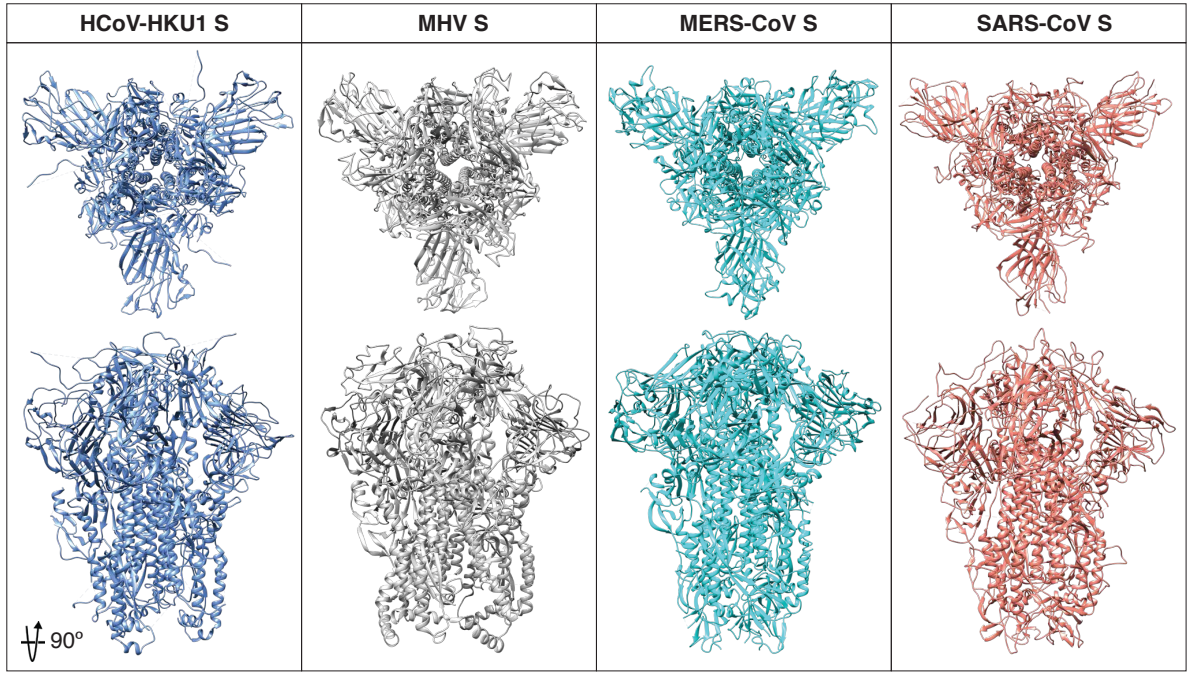

**Supplementary figure 3.** *Betacoronavirus* **S protein structures. A.** Amino acid identity table between Betacoronavirus S proteins. **B.** *Betacoronavirus* S protein structures: HCoV-HKU1, MHV, MERS-CoV and SARS-CoV.

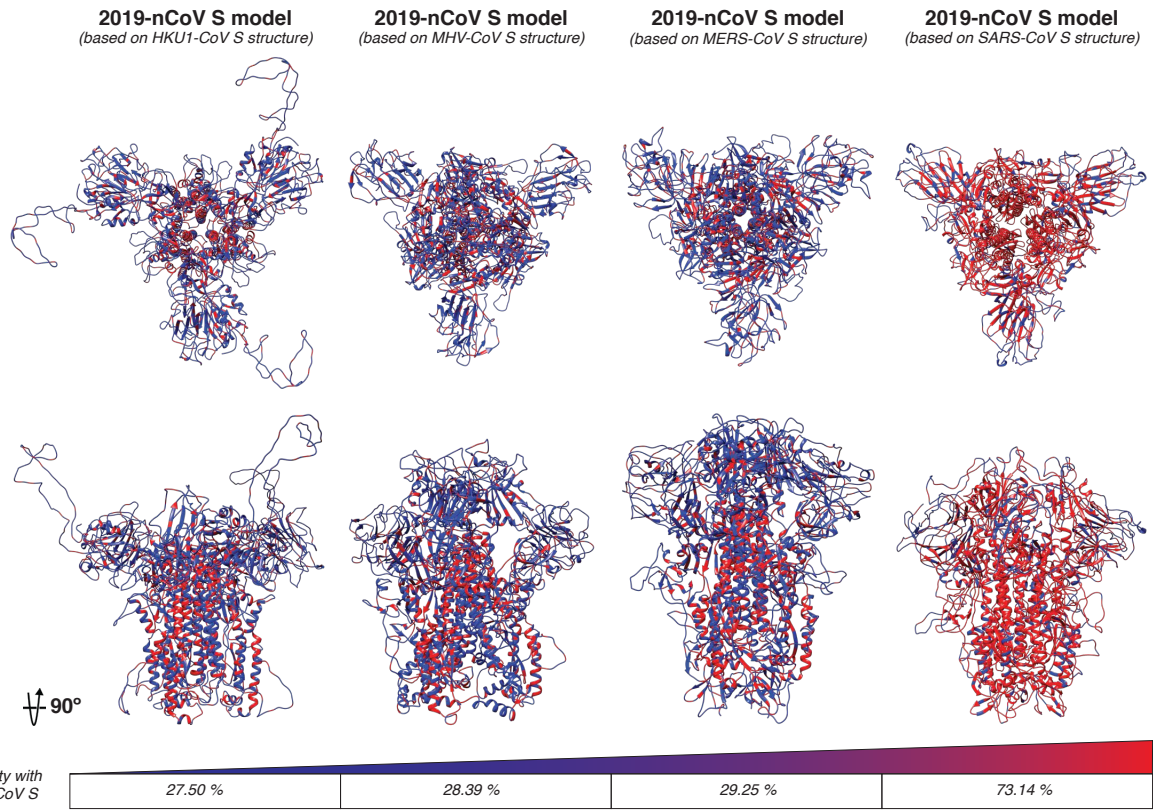

**Supplementary figure 4. 2019-nCoV S protein models based on *Betacoronavirus* structures.** The 2019-nCoV S protein models were built based on the structure of HCoV-HKU1, MHV, MERS-CoV and SARS-CoV. The amino acid homology (red) and differences (blue) are noted. Amino acid identity scale is also shown.